\documentclass{article}
\usepackage{PRIMEarxiv}
\usepackage{fancyhdr}
\usepackage{amsmath,amsfonts,amssymb}
\usepackage{graphicx}
\usepackage[colorlinks=true, allcolors=blue]{hyperref}
\usepackage{booktabs}
\usepackage{caption}
\usepackage{subcaption}
\usepackage{cite}  

\makeatletter
\newcommand{\preprintnotice}{\textrm{This is a preprint of an article published in: Proc. SPIE 13925, Medical Imaging 2026: Image Processing, 139252D (April 2026). https://doi.org/10.1117/12.3086603}}

\providecommand{\@noticestring}{\preprintnotice}

\title{Multimodal synthesis of MRI and tabular data with diffusion in a joint latent space via cross-attention}

\author{
\rlap{Daniel Mensing$^{1,}\thanks{Equal contribution.}$}\hphantom{Daniel Mensing$^{1,\ast}$}, Jan Kapar$^{2,3,\ast}$, Jochen G. Hirsch$^{1}$, Matthias Günther$^{1,4}$, Horst Hahn$^{1,3}$\\
\textbf{Marvin N. Wright$^{2,3}$}
\and
$^1$Fraunhofer Institute for Digital Medicine MEVIS, Bremen, Germany
\and
$^2$Leibniz Institute for Prevention Research and Epidemiology -- BIPS, Bremen, Germany
\and
$^3$Faculty of Mathematics and Computer Science, University of Bremen, Bremen, Germany
\and
$^4$Faculty of Physics and Electrical Engineering, University of Bremen, Bremen, Germany\\
\\
\texttt{daniel.mensing@mevis.fraunhofer.de}
}

\begin{document}

\maketitle
\@notice

\begin{abstract}
We propose a multimodal latent diffusion model that jointly synthesizes volumetric magnetic resonance imaging (MRI) and tabular clinical data within a shared latent space via cross-attention. This approach enables coherent joint representation learning of MRI and tabular modalities for generative modeling. Our model utilizes a variational autoencoder to fuse the two modalities before diffusion-based synthesis, allowing modality-appropriate reconstruction with separate decoders for MRI and tabular data.
We evaluated the framework on data from the German National Cohort (NAKO Gesundheitsstudie), comprising over 10,000 participants with MRI scans and clinical tabular features such as age, sex, body measurements, and ethnicity. The generated MRI volumes exhibited anatomical plausibility and body composition consistent with the synthesized tabular attributes. Quantitative evaluation using Fréchet distance and precision-recall metrics confirmed high-fidelity image generation. In the tabular modality, our model outperformed CTGAN across standard evaluation metrics and achieved results comparable to TVAE, demonstrating competitive performance relative to established unimodal baselines.
This work is, to our knowledge, the first to demonstrate the feasibility of jointly modeling MRI and mixed-type tabular data in a single latent diffusion framework, offering a proof-of-concept for generating coherent synthetic multimodal patient data and aligning with the broader goal of developing digital twins in healthcare.
\end{abstract}

\keywords{
multimodal diffusion, generative modeling, synthetic data, MRI data, tabular data, cross-attention
}

\section{Introduction}
\label{sec:purpose}

Recent advances in generative artificial intelligence (AI), particularly the emergence of denoising diffusion probabilistic models\cite{ho2020DDPM, song2020DDPMScore}, have substantially improved the quality, realism, and controllability of synthetic medical images, including magnetic resonance imaging (MRI). Synthetic MRI data hold great promise for a variety of clinical and machine learning applications, ranging from segmentation to classification tasks \cite{guo2025maisi}. Moreover, synthetic data can help mitigate common challenges in medical imaging---such as limited data availability and patient privacy concerns---by enabling data augmentation to improve model generalization, class balancing in imbalanced cohorts, and privacy-preserving data sharing\cite{shin2018synMRIaugpriv}.

In clinical practice, medical images are routinely combined with complementary data types to form a comprehensive profile of each patient, supporting more accurate diagnosis, prognosis, and treatment planning. Similarly, discriminative AI models for tasks such as classification, anomaly detection, and outcome prediction have been shown to benefit from the inclusion of tabular clinical features---such as demographics, anthropometric measurements, vital signs, laboratory biomarkers, neuropsychological scores, and standardized diagnostic assessments---alongside medical imaging data\cite{krones2025multimodalReview}.

Together, these factors justify the development of multimodal generative models capable of synthesizing coherent pairs of MRI and tabular data and setting the ground for further tasks such as modality-conditional generation\cite{Xin2024Diffcrosscond} and cross-modality imputation\cite{zhang2024crossmodImputation}. However, generative modeling across these heterogeneous modalities remains challenging: deep learning architectures (e.g., diffusion models) excel in generating high-dimensional imaging data, while tree-based models\cite{nowok2016synthpop, watson2023ARF} have historically performed better on mixed-type tabular data\cite{grinsztajn2022treeDLtab, borisov2022treeDLtab, shwartz2022treeDLtab}. As a workaround addressing this heterogeneity, multimodal synthesis can be approached sequentially using modality-specific methods--—for example, by first generating tabular data and then conditionally synthesizing corresponding MRI scans with suitable unimodal approaches. Such pipelines, however, offer limited bidirectionality, and may suffer from impaired inter-modal dependency modeling and error propagation during generation due to exposure bias\cite{schmidt2019ARexposurebias}.

Promisingly, recent diffusion-based models for mixed tabular data have begun to close this heterogeneity gap, demonstrating competitive performance across standard tabular benchmarks \cite{kotelnikov2023tabddpm, zhangTabSyn}. This opens the door to using diffusion models as a unified generative framework across both modalities\cite{Chen2024Diffmultitask}. Yet, to our knowledge, no existing approach jointly synthesizes MRI and tabular data in an end-to-end architecture. In this work, we propose a multimodal generative framework based on diffusion in a shared latent space, where volumetric MRI and tabular clinical data are fused via cross-attention\cite{vaswani2017transformer} within an upstream variational autoencoder (VAE) \cite{kingma2014VAE} to produce a joint latent representation for synthesis. We evaluate this approach using large-scale data from the German National Cohort (NAKO Gesundheitsstudie, NAKO)\cite{peters2022NAKO}, demonstrating its ability to generate anatomically plausible MRIs consistent with patient-specific clinical features through a quantitative analysis based on established imaging and tabular metrics. This multimodal integration represents a step toward high-fidelity digital twins, providing a more holistic representation of patient profiles in synthetic cohorts. 

\section{Methods}
\label{sec:methods}

\subsection{Model \& Training}
The proposed multimodal latent diffusion model (LDM) \cite{rombach2022high} leverages a VAE as its underlying representational bridge. A VAE consists of an encoder-decoder pair that maps high-dimensional data into a lower-dimensional latent manifold. While standalone VAEs are often used as generative models by encouraging the latent space to closely follow a standard Gaussian prior, our approach utilizes the VAE primarily for efficient spatial compression and modality fusion between the image domain \(\mathcal{D}_{img}\) and the tabular domain \(\mathcal{D}_{tab}\). In this approach, we allow the VAE to prioritize high-fidelity reconstruction over a perfectly Gaussian latent structure, relying on the inherent flexibility of the subsequent diffusion model to handle the resulting non-Gaussian distribution.

For tabular inputs, we define the domain as a combination of continuous and discrete spaces $\mathcal{D}_{tab}\ = \mathcal{D}_{num} \times \mathcal{D}_{cat}$. We employed tokenization and separate learnable embedding schemes for numerical features and categorical features, following an approach similar to TabSyn, a state-of-the-art LDM for tabular data \cite{zhangTabSyn}. These embeddings map heterogeneous variables into a unified vector space, enabling the encoder to process them alongside spatially-aware features extracted via 3D convolutions. To reconstruct each modality, distinct decoders were used, enabling modality-appropriate architectures and reconstruction losses. As in the encoding process, 3D convolutions were employed for images and transformers for tabular data.

For image data, we used $\mathcal{L}_1$ loss, a perceptual loss $\mathcal{L}_\text{perc}$ based on intermediate features extracted from a pre-trained network\cite{mmm}, and an adversarial loss $\mathcal{L}_\text{adv}$ from a simple discriminator to assess reconstruction quality similar to the original LDM VAE implementation\cite{rombach2022high}. To measure the reconstruction ability on tabular data, we utilized mean squared error $\mathcal{L}_\text{MSE}$ for numerical and cross-entropy loss $\mathcal{L}_\text{CE}$ for categorical features. Finally, a Kullback–Leibler divergence term $\mathcal{L}_\text{KL}$ was included to regularize the latent space toward a standard normal distribution\cite{kingma2014VAE}. The overall VAE loss was defined as
\begin{equation*}
    \mathcal{L}_\text{VAE} = \underbrace{\mathcal{L}_1 + \lambda_1\mathcal{L}_\text{perc} + \lambda_2\mathcal{L}_\text{adv}}_{\text{Image reconstruction}} \quad + \underbrace{\vphantom{x_p}\mathcal{L}_\text{MSE} + \mathcal{L}_{\text{CE}}}_{\text{Tabular reconstruction}} + \underbrace{\vphantom{x_p}\lambda_3\mathcal{L}_\text{KL}}_{\text{Regularization}},
\end{equation*}
where we set $\lambda_1 = 0.3$, $\lambda_2 = 0.1$, and $\lambda_3 = 10^{-6}$.

Following VAE convergence, we modeled the distribution of the fused latent space using an LDM. By training in this latent space rather than the high-dimensional pixel space, computational complexity is significantly reduced while maintaining generative fidelity. Generally, diffusion models generate data by learning to iteratively reverse a forward process that gradually adds Gaussian noise. As this reverse process is often computationally expensive due to the large number of sequential steps, we employed a denoising diffusion implicit model (DDIM) \cite{song2020ddim}. DDIMs accelerate sampling by employing a non-Markovian trajectory, which enables skipping many intermediate sampling steps while remaining deterministic. We implemented the LDM using a U-Net architecture trained with a linear beta schedule, based on the MONAI framework \cite{cardoso2022monai}.

\subsection{Data}
We used data from the German National Cohort for our experiments. The NAKO is Germany’s largest population-based cohort study, tracking over 200\,000 participants aged 19 to 74 over an extended period. It includes comprehensive data on socioeconomic status, demographic factors, genetics, lifestyle, medical history, clinical examinations, and imaging data\cite{peters2022NAKO}. In NAKO's MR imaging study\cite{bamberg2015NAKO3D}, imaging was performed on MAGNETOM Skyra 3T scanners (Siemens Healthineers, syngo VD13C) using a two-point Dixon volumetric interpolated breath-hold examination (VIBE) T1-weighted sequence. Volumes were acquired in the axial plane with an in-plane matrix of 320 × 260 (spatial resolution: 1.4 × 1.4 mm²) and a slice thickness of 3 mm. Each scan consisted of four table positions, resulting in a total of 316 slices per subject.

For our experiments, we used imaging and tabular data from 10\,748 participants. Specifically, for MRI, we utilized the opposed-phase contrast (TE = 1.23 ms) generated as part of the Dixon VIBE sequence. The MRI volumes were resampled to a size of 160 × 160 × 128 voxels, effectively doubling the voxel size and reducing memory requirements by approximately half to enable GPU processing. Intensity values were normalized to the [0, 1] range. For tabular data, we selected a subset of features available in the NAKO: age, sex, height, weight, body mass index, body fat percentage, and ethnicity. The dataset was split into training and test subsets with a 90:10 ratio.

\section{Results}
\label{sec:results}

\subsection{Synthetic MRI data}

\begin{figure}[t]
    \centering
    \begin{subfigure}{0.45\textwidth}
        \centering
        \includegraphics[width=\linewidth]{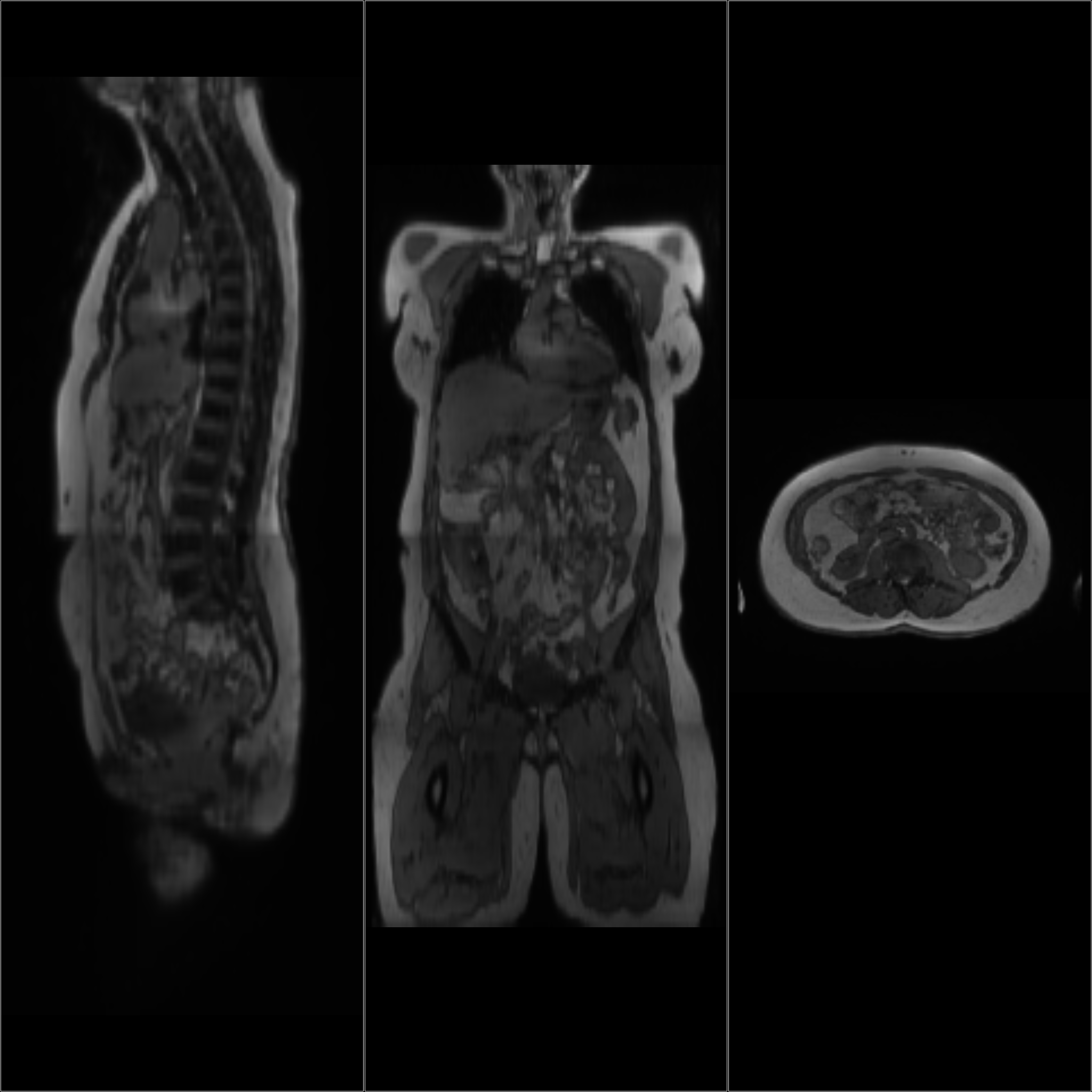}
        \caption{Synthetic MRI volume with the following corresponding synthetic tabular data: age 37.73, sex female, height 165.41, weight 63.90, body mass index 22.79, body fat percentage 37.17, ethnicity European}
        \label{fig:vol1}
    \end{subfigure}
    \hfill
    \begin{subfigure}{0.45\textwidth}
        \centering
        \includegraphics[width=\linewidth]{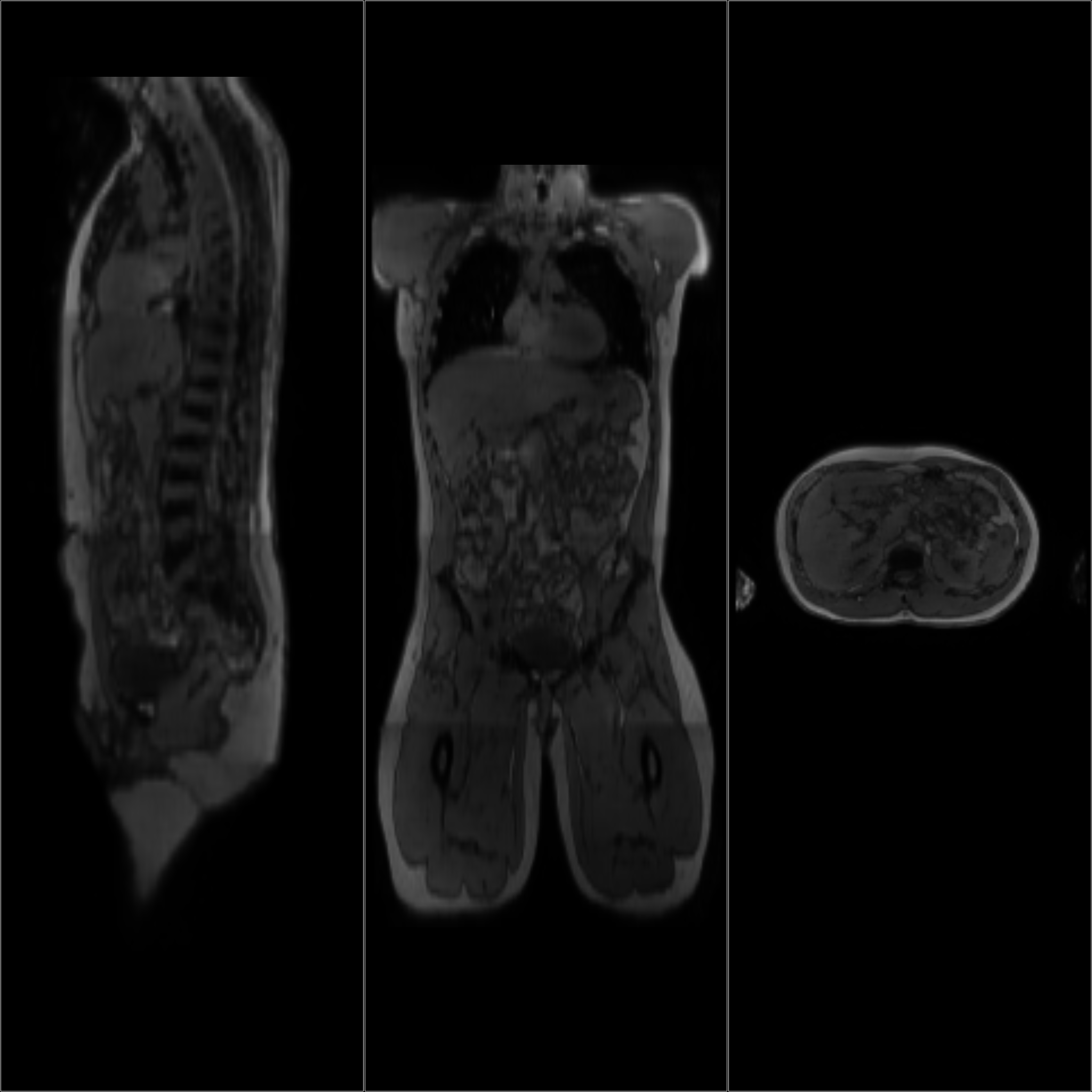}
        \caption{Synthetic MRI volume with the following corresponding synthetic tabular data: age 22.32, sex male, height 196.71,	weight 105.95, body mass index 28.61, body fat percentage 34.75, ethnicity European}
        \label{fig:vol2}
    \end{subfigure}
    \vspace{5pt}
    \caption{Center slices for each orientation for two synthetic MRI volumes with their corresponding synthesized tabular attributes.}
    \label{fig:mri_volume}
\end{figure}

Fig.~\ref{fig:mri_volume} presents example central slices from a synthesized MRI volume for visual inspection, demonstrating anatomical details and body composition consistent with the corresponding tabular variables. To quantitatively assess the synthesis quality, we extracted feature embeddings from both real and generated volumes using the Universal bioMedical PreTrained model (UMedPT) \cite{mmm}. We then computed the Fréchet distance (FD) \cite{heusel2018ganstrainedtimescaleupdate} between these embeddings, obtaining a score of 1.54. We further evaluated image fidelity and distribution coverage via \(\alpha \)-precision (0.871) and \(\beta \)-recall (0.183)\cite{alaa2022alphaprec}. While the \(\beta \)-recall indicates that the model primarily focuses on common anatomical variations within the 10,748-sample dataset, the generated volumes maintain a high level of fidelity, suggesting their potential utility for clinical downstream applications.

\subsection{Synthetic tabular data}

\begin{table}[t]
    \caption{Tabular data synthesis performance comparison of our multimodal LDM against unimodal baselines. LDM, latent diffusion model; WD, multivariate Wasserstein distance (Sinkhorn approximation)\cite{chizat2020WD_Sinkhorn} (range: $(0,\infty))$; Detect., multivariate two-sample detection score\cite{qian2024synthcity} (range: $(0.5,1)$); $\alpha$-Prec., $\alpha$-Precision\cite{alaa2022alphaprec} (range: $(0,1)$), $\beta$-Rec.; $\beta$-Recall\cite{alaa2022alphaprec} (range: $(0,1)$), Univ. Dist., mean distance of univariate distributions\cite{zhangTabSyn} (Kolmogorov-Smirnov distance\cite{Huber1981KS_TV} for numeric features, total variation distance\cite{Huber1981KS_TV} for categorical features) (range: $(0,1)$); Cor. Dist., mean distance of pairwise correlations for mixed data\cite{zhangTabSyn,zhu2022corrdiff} (using the Pearson correlation coefficient $\rho$ for numeric-only pairs, the correlation ratio $\eta$ for mixed pairs, and Cramer's V for categorical-only pairs)\cite{warner2012mixedCor}  (range: $(0,1)$)}
    \label{tab:res_tabular}
    \begin{center}
    \begin{tabular}{lrrrrrr}
        \toprule
        Synthesizer & WD $\downarrow$ & Detect. $\downarrow$ & $\alpha$-Prec. $\uparrow$ & $\beta$-Rec. $\uparrow$ & Univ. Dist. $\downarrow$ & Cor. Dist. $\downarrow$\\
        \midrule
        \textit{Unimodal baselines models:} & & & & & & \\
        \quad TVAE\cite{Xu2019CTGAN} & 0.306 & 0.905 & 0.260 & 0.134 & 0.093 & 0.060 \\ 
        \quad CTGAN\cite{Xu2019CTGAN} & 0.809 & 0.952 & 0.127 & 0.037 & 0.120 & 0.130 \\ 
        \quad TabSyn\cite{zhangTabSyn} & 0.251 & 0.513 & 0.944 & 0.412 & 0.007 & 0.007 \\ 
        \midrule
        Multimodal LDM (ours) & 0.470 & 0.934 & 0.562 & 0.168 & 0.139 & 0.035 \\ 
        \bottomrule
    \end{tabular}
    \end{center}
\end{table}

For the tabular modality, we compare the synthesis quality of our multimodal LDM against three established unimodal baselines: TabSyn \cite{zhangTabSyn}, a state-of-the-art diffusion-based model for tabular synthesis, as well as CTGAN and TVAE \cite{Xu2019CTGAN}, which remain widely utilized benchmarks. Notably, TabSyn serves as a unimodal reference for our approach, as it focuses exclusively on the tabular domain. Our framework, however, must learn the complex dependencies required to maintain consistency within a high-dimensional joint MRI-tabular space.

To assess synthesis quality comprehensively, we applied metrics spanning multivariate, bivariate, and univariate aspects of the data. These included the multivariate Wasserstein distance \cite{chizat2020WD_Sinkhorn} and synthetic data detection \cite{qian2024synthcity} for overall distribution similarity, \(\alpha \)-precision and \(\beta \)-recall for sample fidelity and coverage \cite{alaa2022alphaprec}, as well as the mean univariate distance and mean absolute error of pairwise correlations \cite{zhangTabSyn}.
Table~\ref{tab:res_tabular} summarizes these results. As expected, TabSyn achieves the best scores across all metrics. Our multimodal LDM demonstrates competitive performance relative to CTGAN and TVAE, outperforming CTGAN in all metrics except for the mean univariate fit. Notably, the low correlation distance achieved by our model suggests a superior ability to capture dependencies within the data compared to the GAN-based baselines. This is further supported by our model's \(\alpha \)-precision, which substantially exceeds that of CTGAN and TVAE. While the \(\beta \)-recall remains relatively low, it still represents an improvement over these baselines, indicating better overall distribution coverage. Further refinement of univariate modeling remains an objective for future work.

\section{Conclusion}
\label{sec:conclusion}

In this work, we presented a multimodal latent diffusion model that jointly synthesizes volumetric MRI and tabular clinical data by fusing both modalities in a shared latent space via cross-attention. Our experiments demonstrate the feasibility of this joint approach, achieving compelling performance in generating anatomically plausible MRI volumes and outperforming established baselines, such as CTGAN and TVAE, across several key metrics for tabular synthesis. This framework represents a novel departure from common sequential and inherently unidirectional pipelines by implementing a joint generative strategy for these heterogeneous modalities.

While these results are promising, further refinements remain an objective to bridge the performance gap with specialized unimodal models. In particular, there is potential to enhance distribution coverage and diversity across both modalities, as well as to improve the modeling of univariate tabular distributions. Future research will focus on optimizing model architectures and training strategies to tackle these challenges and enhance overall generative quality. Furthermore, we aim to expand the clinical feature set to evaluate more complex cross-modality dependencies and to investigate the inherent advantages of joint synthesis over conditional LDM variants.

Moving forward, we plan to build upon this foundation by integrating unified transformer architectures \cite{bao2023diffmultimodalTrans, rojas2025diffuseEverything}, which have shown potential for applications such as cross-modality imputation and modality-conditional generation. By enabling such bidirectional workflows, this framework represents a significant step toward the development of comprehensive medical digital twins, thereby informing clinical decision-making and patient care.

\section*{Acknowledgements} 
This work was supported by the U Bremen Research Alliance / AI Center for Health Care, funded by the Federal State of Bremen. We used data (application numbers NAKO-246 and NAKO-839) from the German National Cohort (NAKO Gesundheitsstudie, NAKO) (www.nako.de). Scientists can apply for data access following the official usage regulations and upon formal request to the NAKO use and access committee (https://transfer.nako.de). The NAKO is funded by the Federal Ministry of Education and Research (BMBF) [project funding reference numbers: 01ER1301A\slash B\slash C, 01ER1511D, 01ER1801A\slash B\slash C\slash D and 01ER2301A\slash B\slash C], federal states of Germany and the Helmholtz Association, the participating universities and the institutes of the Leibniz Association. We thank all participants who took part in the NAKO study and the staff of this research initiative.

\bibliographystyle{unsrt}
\bibliography{main.bib}

\end{document}